\documentstyle[12pt,epsfig]{article}
\begin{document}

\newcommand{\ccaption}[2]{
    \begin{center}
    \parbox{0.85\textwidth}{
      \caption[#1]{\small{{#2}}}
      }
    \end{center}
    }

\thispagestyle{empty}
\vspace*{2cm}
  \begin{Large}\begin{center}
{\sf \sf $\rho_T$ Production via $W_L$ $Z_L$ Fusion at Hadronic Colliders }
\end{center}
\end{Large}

\normalsize
\vspace*{0.5cm}
\begin{center}
S.R.~Slabospitsky\\
{\it Institute for High Energy Physics, \\ Protvino, Moscow Region,
142284 Russia}\\
\verb+ slabospitsky@mx.ihep.su+\\
\bigskip
G.~Azuelos \\ 
{\it  Universit\'e de Montreal, Canada} \\
\verb+ azuelos@lps.Umontreal.ca+\\ 
\end{center}                  

\nopagebreak
\vfill
\vskip 3cm
\begin{abstract} 
Multiscale technicolor models predict the existence of high mass resonances
at hadron colliders. Although the quark fusion process of production
dominates, vector boson fusion offers the advantage of allowing
forward jet tagging for background suppression. 
We calculate here the cross section and differential distributions for 
$\rho_T$ production in the vector boson fusion channel at the LHC.
\end{abstract}                                                          

\vskip 1cm
%\today \hfill 
\vfill 

\newpage
\setcounter{page}{1}

\section{\bf Introduction }
   
  In the search for the origin of electroweak symmetry breaking, 
strong coupling models are an alternative to the Standard Model
Higgs mechanism, or for weakly coupled theories based on supersymmetry.
Although classical technicolor\cite{technicolor} suffers from major
shortcomings, viable multiscale models
have been developed~\cite{Lane} which allow for fermion mass generation
and for the absence of weak neutral currents.  These models, which are
not necessarily excluded by precision measurements of electroweak parameters 
at LEP and SLC~\cite{ewconstraints}, not only 
provide the technipions which serve as the longitudinal components of
the gauge bosons, but also predict  
the existence of technipions ($\pi_T$) as mass eigenstates, 
as well as technirhos ($\rho_T$), and
other techniparticles.
 
 The feasibility of observing such resonances produced in $q\bar q$ fusion, in
 ATLAS has been reported in~\cite{TDR}. The analysis was based on a 
PYTHIA~\cite{PYTHIA} implementation of the multiscale model of Eichten, Lane 
and Womersley~\cite{Lane}, and took into
account detector effects by using a fast
simulation of the ATLAS detector. In particular, 
one of the channels analyzed was $\rho_T^\pm$
production with subsequent decay into 
$\pi_T^0 W_L^\pm \to b \bar b \ell^\pm \nu$.
Assuming a quark fusion process, the authors~\cite{Lane} of the model have 
calculated the relevant matrix elements. It has then been shown in~\cite{TDR} 
that, given some reasonable values for the parameters of the model,
the observation of such a process should be feasible with ATLAS,
but is limited by background for possible large masses of $\rho_T$ and
$\pi_T$. 

 The technique of forward jet tagging has been shown~\cite{TDR} 
to be very powerful in
eliminating backgrounds from such processes as $W$ + jets production.
For that reason, it is important to estimate if $\rho_T$ production via 
a vector boson fusion process can
be a useful complementary channel for discovery of such a resonance. We
evaluate here the cross section for this process.

The article is organized as follows. First, the model for the 
form factor and decay width of $\rho_T$ are 
presented in section 2. Results on the cross section calculation are
given in section 3.  
The main results are summarized in the conclusion.

\section {\bf Model }

In this section we describe in detail the essential features of the model 
used
in our calculations. 

The lighter isotriplet $\rho_T$ is assumed to decay 
dominantly into pairs of 
the mixed state of isotriplet 
$|\Pi_T \rangle = \sin \chi |W_L \rangle + \cos \chi |\pi_T\rangle$,
where the value of the mixing angle $\chi$ is  assumed~\cite{Lane} to  
be: $\sin \chi = 1/3$.
The vertex $\rho_T \to \pi_A \pi_B$, where $\pi_{A,B}$ may be longitudinal
$W^{\pm}_L, Z_L$ bosons or technipion $\pi_T$, has the 
following form~\cite{Lane}:
\begin{eqnarray}
 g_{\rho} F_{\rho \pi \pi}(p, q_A, q_B) C_{AB} \varepsilon^{\nu} 
(q_A - q_B)_{\nu} 
 \label{vrt1}
\end{eqnarray}
where $p$, $q_A$ and $q_B$ are the momenta of $\rho_T$, $\pi_A$ and $\pi_B$;
$\varepsilon^{\nu}$ is the polarization vector of $\rho_T$;
the parameters $C_{AB}$  are equal to 
\begin{eqnarray}
 C_{AB} = \left\{  
\begin{array}{lcl} 
\sin^2 \chi & {\rm for} & W^{\pm}_L \; Z_L, \\
       \sin \chi \cos \chi& {\rm for} & W^{\pm}_L \; \pi_T, \\
       \cos^2 \chi & {\rm for} & \pi_T \; \pi_T 
\end{array} \right.   \label{cab}
\end{eqnarray}
and $g_{\rho}$ is a coupling constant, normalized as follows~\cite{Lane}:
\begin{eqnarray*}
\frac{ g_{\rho}^2}{4 \pi} = 2.91 \left( \frac {3} {N_{TC}} \right), 
\end{eqnarray*}
where $N_{TC} = 4$ (see~\cite{Lane}).

  Being a compound object, consisting of two heavy techniquarks,
the technirho couples to two technipions (or longitudinal $W, Z$ bosons)
by means of the diagram in Fig.1. In the loop the techniquarks 1 and 2
are on-shell (it is a typical approximation for a such consideration), while
quark~3 is virtual. The latter has a momentum $p_3$ given by:
\[
p_3 = p_1 - q_A = \frac{1}{2}p_{\rho} - q_A = \frac{1}{2}(q_B - q_A).
\]

% Figure 1
\begin{figure}[htb]
\begin{center}
\epsfig{file=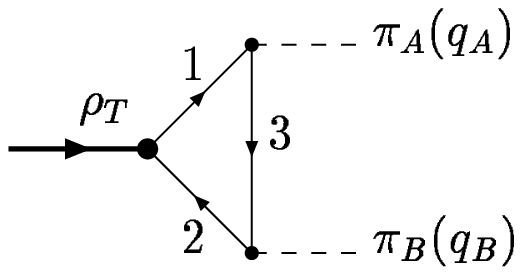,width=8cm} 
\ccaption{}{
Diagram describing $\rho_t \pi_A \pi_B$ coupling via a loop of heavy
techniquarks (lines 1, 2, and 3). The quarks $Q_1$ and $Q_2$ are on-shell
($p_{Q_1} = -p_{Q_2} = \frac{1}{2} p_{\rho}$, and the momentum of the 
off-shell quark $Q_3$ is $\frac{1}{2}(q_B - q_A)$.
}
\end{center}
\end{figure}

The production vertex
 is naturally suppressed by large virtualities of the $W_L$ and
$Z_L$, but is enhanced, in that case, by large values of 
$p_3$ in the numerator. 
To be consistent
with the Effective W Approximation (EWA), the additional form
factor $ F_{\rho \pi \pi}(p, q_A, q_B)$ in (\ref{vrt1}) is introduced.

\begin{eqnarray}
F_{\rho \pi \pi}(p, q_A, q_B) = \frac{ M^2_{\rho} - M^2_{\pi_A} - M^2_{\pi_B}}
 {2 (q_A q_B)}, \label{ff1} 
\end{eqnarray}
where $M_{\rho}, M_{\pi_{A,B}}$ are the masses of $\rho_T, \pi_{A,B}$. 
In analogy with the case of a
heavy $(Q_1 \bar Q_2)$-meson, this form-factor takes into 
account the possible off-shellness of technipions
$\pi_A$ and $\pi_B$. 
The denominator 
in the effective $\rho_T \pi_A \pi_B$ vertex (\ref{ff1}) results from
the propagator of this virtual quark in the loop:
\begin{eqnarray}
 m_3^2 - p_3^2 = \frac{M^2_{\rho}}{4} - \frac{(q_B - q_A)^2}{4} = (q_A q_B),
 \label{den1} 
\end{eqnarray}
where $ M^2_{\rho} = q^2_A + q^2_B + 2 (q_A q_B)$. 

 In the decay
vertex of $\rho_T$, for on-shell technipions $\pi_A$ and 
$\pi_B$, we have $F_{\rho \pi \pi}|_{\rm on-shell} = 1$. 

Using the vertex (\ref{vrt1}) we get the well-known equation for the 
$\rho_T \to \pi_A \pi_B$ decay width~\cite{Lane}:
\begin{eqnarray}
 \Gamma(\rho_T \to \pi_A \pi_B) = \frac{ 2 \alpha_{\rho_T} C_{AB}^2}{3}
 \frac{p^3_{AB}}{M^2_{\rho}}  \label{gamro} 
\end{eqnarray}

We investigate the case of $\rho^{\pm}_T$ production with its subsequent 
decay 
into $\pi^0_T W^{\pm}_L$ pair. The corresponding branching ratio has a
non-trivial behavior, as can be seen in Fig.2 for 
three decay channels of the charged technirho: 
$\rho^{\pm}_T \to 
\pi^0_T \pi^{\pm}_T$, $\rho^{\pm}_T \to \pi^0_T W_L + \pi^{\pm}_T Z_L$ or 
$\rho^{\pm}_T \to Z_L W^{\pm} _L$. 

% Figure 2
\begin{figure}[htb]
\begin{center}
\epsfig{file=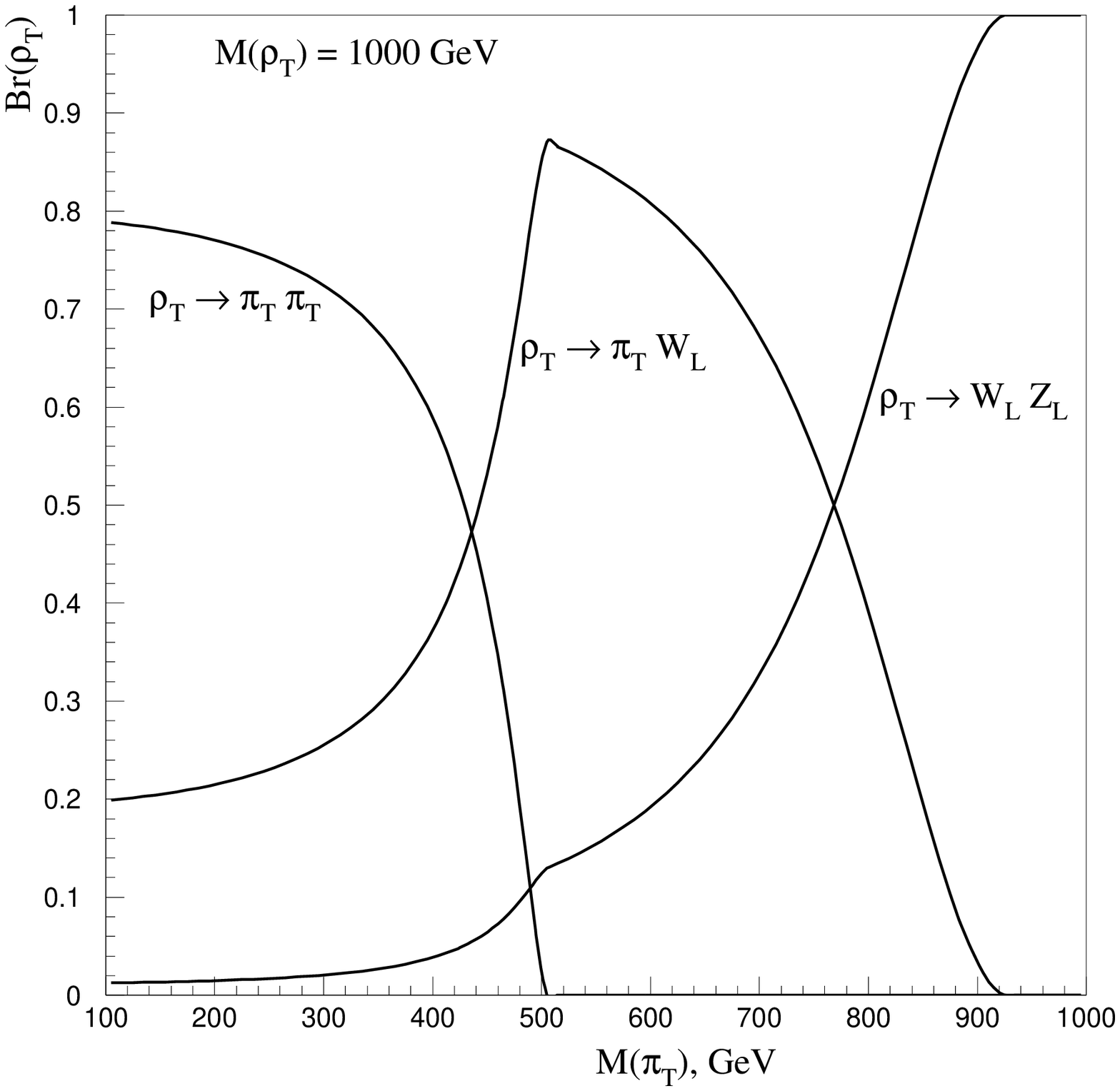,width=8cm,clip=}
\ccaption{}{
Branching ratios versus $M_{\pi_T}$ for the decay channels of 
$\rho^{\pm}_T \to 
\pi^0_T \pi^{\pm}_T$, $\rho^{\pm}_T \to \pi^0_T W_L + \pi^{\pm}_T Z_L$ or 
$\rho^{\pm}_T \to Z_L W^{\pm} _L$. 
We consider the case  $m_{\rho_T}= 1$ TeV and 
$m_{\pi_T}=500$ GeV.
}
\end{center}
\end{figure}

We can distinguish two regions in the technipion to technirho mass 
ratio. In the first region, namely  $M_{\pi_T} / M_{\rho_T} \le  1/2$,
the $\rho^{\pm}_T \to \pi_T \pi_T$ decay is kinematically allowed.
As $M_{\pi_T}$ increases (keeping $M_{\rho_T}$ fixed)
the relative momentum 
% $p_{\pi_T \pi_T}$ (see equation~(\ref{gamro}))  becomes 
 $p_{AB}$ in equation~(\ref{gamro}) decreases.
As a result we have a decreasing branching fraction for the 
$\pi_T \pi_T$ channel,
while the other two channels increase in their relative value (see Fig.~2).
Just above the value $M_{\pi_T} / M_{\rho_T} = 1/2$ 
the branching ratio for the $\pi_T W_L$ channel reaches its maximum value. 
As the mass of $\pi_T$ rises further, in the second region up to
the kinematic bound
$M_{\pi_T} / M_{\rho_T} = 1$, the relative
boson momentum $p_{W \pi_T}$ decreases, yielding a decreasing 
branching ratio for technipion~$+$~longitudinal $W$ boson.

\section {\bf Calculation of the cross section }

We examine the reaction of technirho production at $\sqrt{s} = 14$~TeV (LHC 
collider) with subsequent decay of the technirho into a neutral technipion 
$\pi^0_T$ and a longitudinal $W^{\pm}_L$-boson 
\begin{eqnarray}
p p \to q_f q_f \rho_T \, (\to \pi^0_T W_L )\, X \label{reac1}
\end{eqnarray}
Since each of the final particles from the $\rho_T$ decay has a very 
narrow width in comparison with its mass,  we take both the final
$W_L$-boson and technipion to be on-shell.

One of the diagrams describing the  subprocess 
\begin{eqnarray}
q q' \to q_f q'_f \rho_T \to q_f q'_f \pi^0_T W_L \label{reac3}
\end{eqnarray}
is shown in Fig.~3.
 Only fusion of longitudinal $W_L$ and $Z_L$ 
bosons, radiated from the initial quarks $q$ and $q'$, 
needs to be taken into account.

% Figure 3
\begin{figure}[htb]
\begin{center}
\epsfig{file=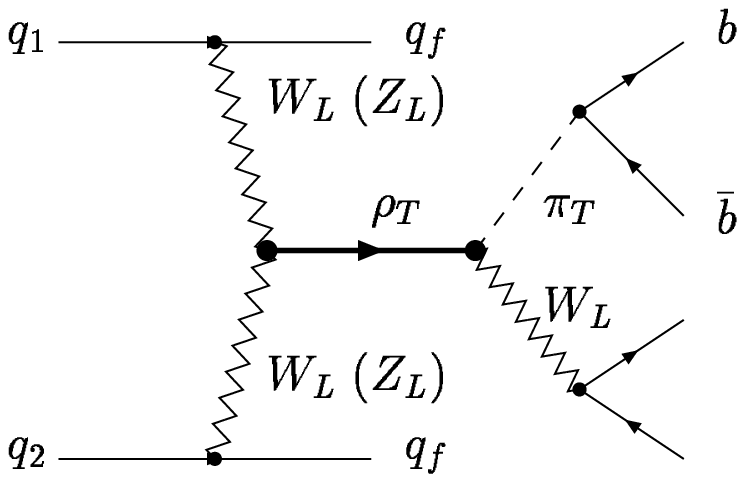,width=8cm} 
\ccaption{}{
One of the diagrams describing $\rho^{\pm}_T$ production with subsequent decay
into $\pi^0_T (\to b \bar b)$ and $W_L^{\pm} (\to l^{\pm} \nu)$.
}
\end{center}
\end{figure}

The polarization vector of a longitudinal boson with 
mass~$M$ and 4-momentum
$q$ is equal to~\cite{wl};
\begin{eqnarray}
 e^{\nu} = \frac{q^{\nu}}{M} + \frac{(q_0 - |\vec q|)}{M} 
(-1, \vec q / |\vec q|). \label{pol1} 
\end{eqnarray}

In our calculations we set that 
\begin{eqnarray*} 
       && {\rm BR}(\pi^0_T \to b \bar b) = 1, \\
{\rm and} && {\rm BR}(W^{\pm}_L \to l^{\pm} \nu) = 1/9.
\end{eqnarray*} 
We use the CTEQ2L parameterization of the proton structure 
functions~\cite{cteq}. For evolution scale we set $\sqrt{Q^2} = M_W$. 

The resulting cross sections for different values of the masses of 
$\rho^{\pm}_T$ and $\pi^0_T$ are presented in the Table~\ref{tab:sig}. 
Note that the cross section of the reaction~(\ref{reac1}) may be presented
in the factorized form as follows:
\begin{eqnarray}
\sigma(p p \to q_f q_f \rho_T (\to \pi^0_T W_L) X)= 
\sigma(p p \to q_f q_f \rho_T X) \times {\rm BR}(\rho_T \to \pi^0_T W_L).
\label{sig1} 
\end{eqnarray}
Therefore, the $M_{\pi_T}$-dependence of the cross section for 
reaction~(\ref{reac1}) with a given value of $M_{\rho_T}$ is completely 
determined
by the branching ratio of technirho decay into $\pi_T W_L$ (see~Fig.2).
In particular, we expect that the cross section of reaction~(\ref{reac1})
should reach a maximum for $M_{\pi_T}/M_{\rho_T} \approx 1/2$. 

A brief remark about the role of the form-factor 
$F_{\rho \pi \pi}$ in the $\rho_T \pi_A \pi_B$ vertex~(\ref{vrt1})
is in order. In 
Fig.4  comparison is made of the pseudorapidity distribution 
of the final primary light quarks 
$q_f, q^{'}_f$ (not from $\pi^0_T$ decay). The dashed histogram corresponds to 
the $q_W$ and
$q_Z$ dependent form-factor $F_{\rho \pi \pi}$ using expression ~(\ref{ff1}), 
while
the solid histogram is produced with $F_{\rho \pi \pi} = 1$. From this figure
one clearly sees that the latter choice ($F=1$) disagrees with 
expectations from the
effective $W$-approximation, since the pseudorapidity
distribution of the final $q_f$ do not peak in the forward 
and backward direction).

% Figure 4
\begin{figure}[htb]
\begin{center}
\epsfig{file=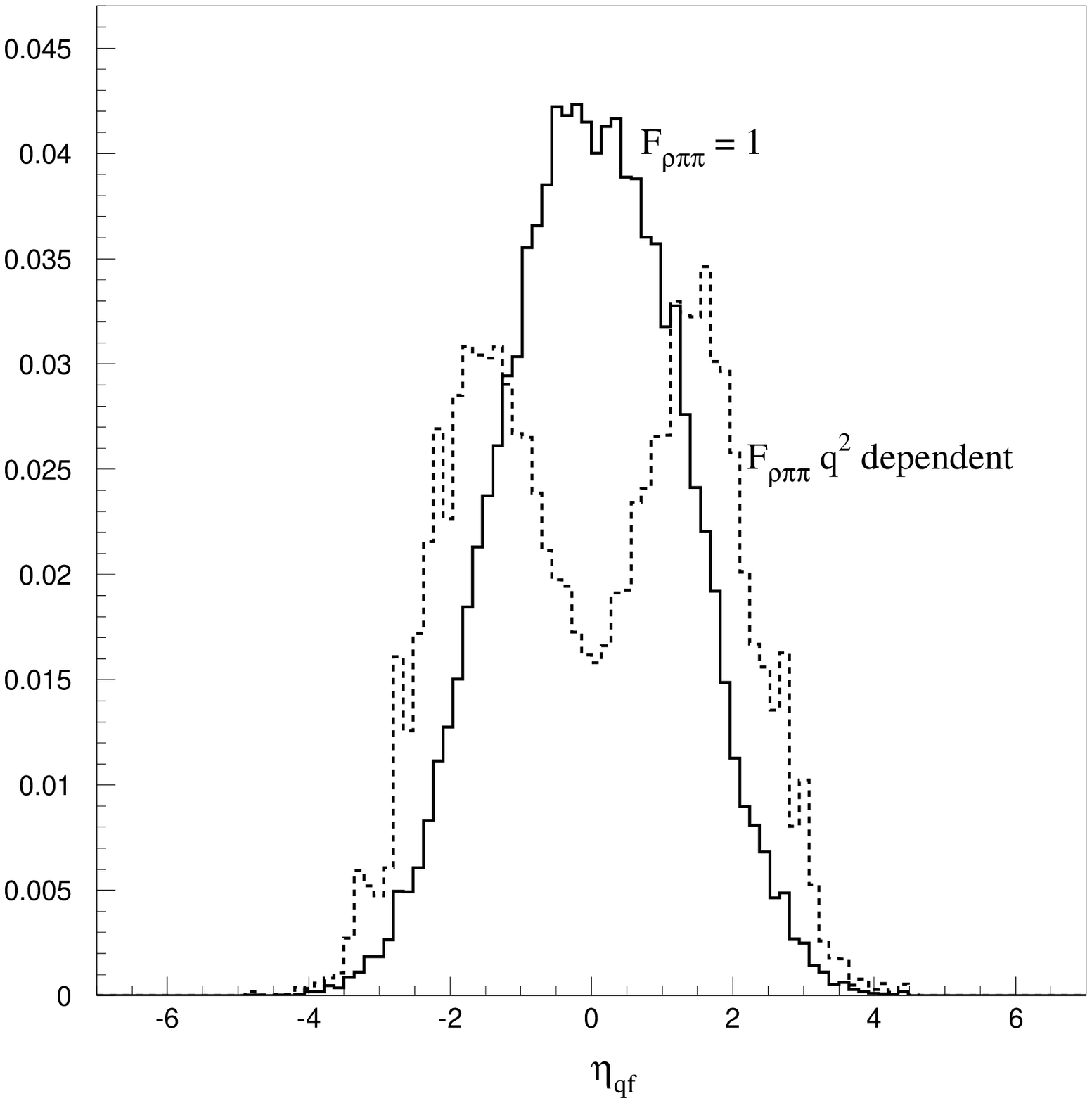,width=16cm} 
\ccaption{}{
Pseudorapidity distribution of final $q_f$ quarks, calculated with
momentum-transfer dependent form-factor $F_{\rho \pi \pi}$ of the 
form of equation~(\ref{ff1}) (dashed histogram) and 
with $F_{\rho \pi \pi} = 1$ (solid
histogram).
}
\end{center}
\end{figure}

Other differential distributions are shown in the Figs. 5-6 (solid histograms).

We also compare our exact results with the calculations within effective 
$W$-approximation approach~\cite{wl}. In this case the matrix element for the
process
\[ W_L \, Z_L \, \to \rho_T \, \to \pi_T \, W_L
\]
looks very simple. The exact form of the $|M|^2$ is presented below 
\begin{eqnarray*}
\nonumber |M|^2 &\propto&
 C_{WZ}^2 C_{\pi W}^2 \left( \frac{(p_w^2 - p_z^2)
(p_a^2 - p_a^2)}{M_{\rho_T}^2} - (p_w - p_z)(p_a - p_b) \right) ^2 \\
 && \times \frac{1}{(P_{\rho_T}^2 - M_{\rho_T}^2)^2 + 
(\Gamma_{\rho_T} M_{\rho_T})^2},
\end{eqnarray*}
where $p_w$ and $p_z$ are the momenta of the initial $W_L$ and $Z_L$, 
while $p_a$ and $p_b$ are the momenta of the final longitudinal $W_L$ boson
and techipion $\pi_T$, respectively (see~\cite{pirs}, for detail). The
parameters $C_{WZ}$ and $C_{\pi W}$ are given by the equation~(\ref{cab}). 

The corresponding differential distributions are shown in the Figs. 5-6 (dashed
histograms). One can see from these figures that the shapes from the exact 
and EWA calculations are in good agreement.

% \section{Observability of $q q \to q q \rho_T^\pm  \to q q W^\pm\pi_T^0$}
% %%%%%%%%%%%%%%%%%%%%%%%%%%%%%%%%%%%%%%%%%%%%%%%%%%%%%%%%%%%%%%%%%%%%%%%%%%

 As an example for the above calculation, the same case as in~\cite{TDR}
but by the process of gauge boson fusion, has been reported
in~\cite{TDR2}: $ \rho_T^\pm   \to q q W^\pm\pi_T^0 \to l\nu b \bar b $
(where $l$ is an electron or muon), with $m_{\rho_T}$ = 800 GeV 
and $m_{\pi_T}$ = 500 GeV. For this process, 
$\sigma\times BR$ is about 2.2 fb, as can be seen from table 1 (accounting
for both $\mu \nu$ and $e \nu$ decays of the charged $W$. 
This cross section depends sensitively on 
the assumed value of the mixing  $\sin \chi$ between the longitudinal gauge 
bosons and the technipions since it involves the 
$WZ\rho_T$ vertex as well as the
$\rho_T W \pi_T$ vertex. It was shown that, with the same model parameters,
 the resulting signal would leave 
about 2.6 events on a background of about 5.6 for an integrated luminosity of 
30 fb$^{-1}$, or a value of $\sigma/\sqrt{B} = 1.1$. This is to be compared 
with a significance of 2.1 obtained in ref~\cite{TDR} for the $q\bar q$
fusion contribution.
This process of vector boson fusion with forward tagging of 
jets could therefore complement
the $q\bar q$ fusion process, but would not be a discovery channel unless the
$\sigma\times BR$ is at least 12 fb.

\section{\bf Conclusion }

We have calculated the cross section and differential distributions for 
$\rho_T$ production, in the vector boson fusion channel. We have compared
our exact calculations with those within the effective W-boson approximation. 
The shapes of the
differential distributions for the final particles (charged leptons and 
$b$-quarks) obtained by the two
methods (the exact and EWA)  are in reasonable agreement. 

For a particular choice of masses for the technirho and technipion
($M_{\rho_T} = 800$~GeV and $M_{\pi_T} = 500$~Gev) 
the possibility of extracting the signal from the
background was evaluated. After applications of
kinematical cuts, assuming low luminosity, 
one would expect about 2.6 events from the signal
and about 5.6 ones from the background. Therefore, the 
process of vector boson fusion with forward tagging of jets could complement
the $q\bar q$ fusion process.

\section* {\bf Acknowledgments}

We thank to D.~Froidevaux, R.~Mazini, A.~Miagkov, V.~Obraztsov, and A.~Zaitsev 
for  fruitful  discussions. S.S. acknowledges the hospitality of Theory
Division of CERN, where this work had complete.

\newpage

\begin{table}[htbp]
%\begin{table}

\caption {}{  The value of the $\rho_T \to \pi^0_T W^{\pm}_L$ production 
cross sections (times branching ratios for $\pi^0_T \to b \bar b$ and 
$W^{\pm}_L \to \mu^{\pm} \nu$) for different values of $M_{\rho_T}$ and 
$M_{\pi_T}$. The masses are in GeV, while the cross sections are in~fb. }
\label{tab:sig}

\begin{center}
\vspace {0.5cm}

\begin{tabular}{|r|c|c|c|c|c|c|c|c|} \hline 
$p p \to \rho^+_T X$ & $M_{\pi} = 200$ 
 & 400 & 500 & 600 & 800 & 1000 & 1200 & 1400     \\ \hline 
$M_{\rho} = 400$
      & 1.924 &       &       &       &       &       &       &       \\ \hline
  500 & 0.630 & 0.268 &       &       &       &       &       &       \\ \hline
  600 & 0.415 & 1.213 & 0.146 &       &       &       &       &       \\ \hline
  800 & 0.247 & 0.752 & 0.690 & 0.610 &       &       &       &       \\ \hline
 1000 & 0.107 & 0.135 & 0.307 & 0.296 & 0.163 &       &       &       \\ \hline
 1200 & 0.088 & 0.079 & 0.079 & 0.281 & 0.172 & 0.077 &       &       \\ \hline
 1500 & 0.105 & 0.057 & 0.038 & 0.075 & 0.079 & 0.089 & 0.038 & 0.001 \\ \hline
 2000 & 0.030 & 0.014 & 0.011 & 0.011 & 0.010 & 0.027 & 0.027 & 0.032 \\ \hline
\hline
$p p \to \rho^-_T X$ & $M_{\pi} = 200$ 
& 400 & 500 & 600 & 800 & 1000 & 1200 & 1400     \\ \hline 
$M_{\rho} = 400$ 
    & 1.288 &       &       &       &       &       &       &       \\ \hline  
500 & 0.391 & 0.166 &       &       &       &       &       &       \\ \hline  
600 & 0.243 & 0.699 & 0.088 &       &       &       &       &       \\ \hline  
800 & 0.129 & 0.399 & 0.378 & 0.319 &       &       &       &       \\ \hline  
1000 & 0.060 & 0.072 & 0.162 & 0.155 & 0.087 &       &       &      \\ \hline  
1200 & 0.050 & 0.040 & 0.042 & 0.148 & 0.087 & 0.038 &       &      \\ \hline  
1500 & 0.060 & 0.030 & 0.019 & 0.036 & 0.038 & 0.042 & 0.019 & 0.001 \\ \hline
2000 & 0.018 & 0.008 & 0.005 & 0.006 & 0.005 & 0.012 & 0.012 & 0.014 \\ \hline
\end{tabular}
\end{center}
\end{table}

\begin{table}
\caption{}{
Signals and  backgrounds  for the observation  of $\rho_T^\pm
\to W^\pm \pi_T^0 \to \ell^\pm \nu b \bar b$. The last column shows
the multiplicative factor used in the normalization of the background,
assuming 30 fb$^{-1}$.}
\label{tab:MC}

\begin{center}
\begin{tabular}{|l|c|c|c|c|} \hline
 Process & preselection & $\sigma$ & Nb. of events        & factor  \\
         &              &  (pb)    & simulated & 30 fb$^{-1}$ \\  \hline
$\rho_T^{\pm}$ &$\hat{m}>600$ GeV/c$^2$ &0.0022 & 2000              & 0.033 \\
$t\bar t$      &$\hat{p}_T>100$  GeV/c  & 333   & $5.3\times 10^6$  & 1.88 \\
W + jets &$W \to \ell \nu$;$\hat{p}_T>100$ GeV/c & 638 &$6 \times 10^6$& 3.2 \\
$W b \bar b$ &   & 19.9           &  $2 \times 10^5$       &  3.0 \\
Z + jets&$Z \to \ell^+\ell^-$; $\hat{p}_T>100$ GeV/c& 85.5 &$6\times 10^5$ 
 & 4.28\\
  WZ     &      &   20.6         & $        10^6$         & 0.618 \\ \hline
\end{tabular}
\end{center}
\end{table}

\begin{table}
\caption{}{
Number of signal/$t\bar t$/(W+jets+Z+jets)  events around the
mass peak of the signal, after the application of successive cuts (see
the text).  The last two lines  give the $\sigma\times BR$  predicted by the
model, as well  as  the
$\sigma\times BR$ required for a 5$\sigma$ significance of the signal}
\label{tab:results}

\begin{center}
\begin{tabular}{|c|c|c|} \hline
 Cut          & $\rho_T$        \\ \hline 
  A           & 6.2/3150/34.7   \\
  B           & 6.0/1400/29.7   \\
  C           & 2.6/5.6/0       \\
 $ S/\sqrt{B}$ &  1.1          \\ \hline
 $\sigma\times BR$  (fb)  &      \\
  model       & 2.5             \\
  for $5\sigma$ significance & 11.6 \\ \hline
\end{tabular}
\end{center}
\end{table}

% Figure 5
\begin{figure}[t]
\begin{center}
\epsfig{file=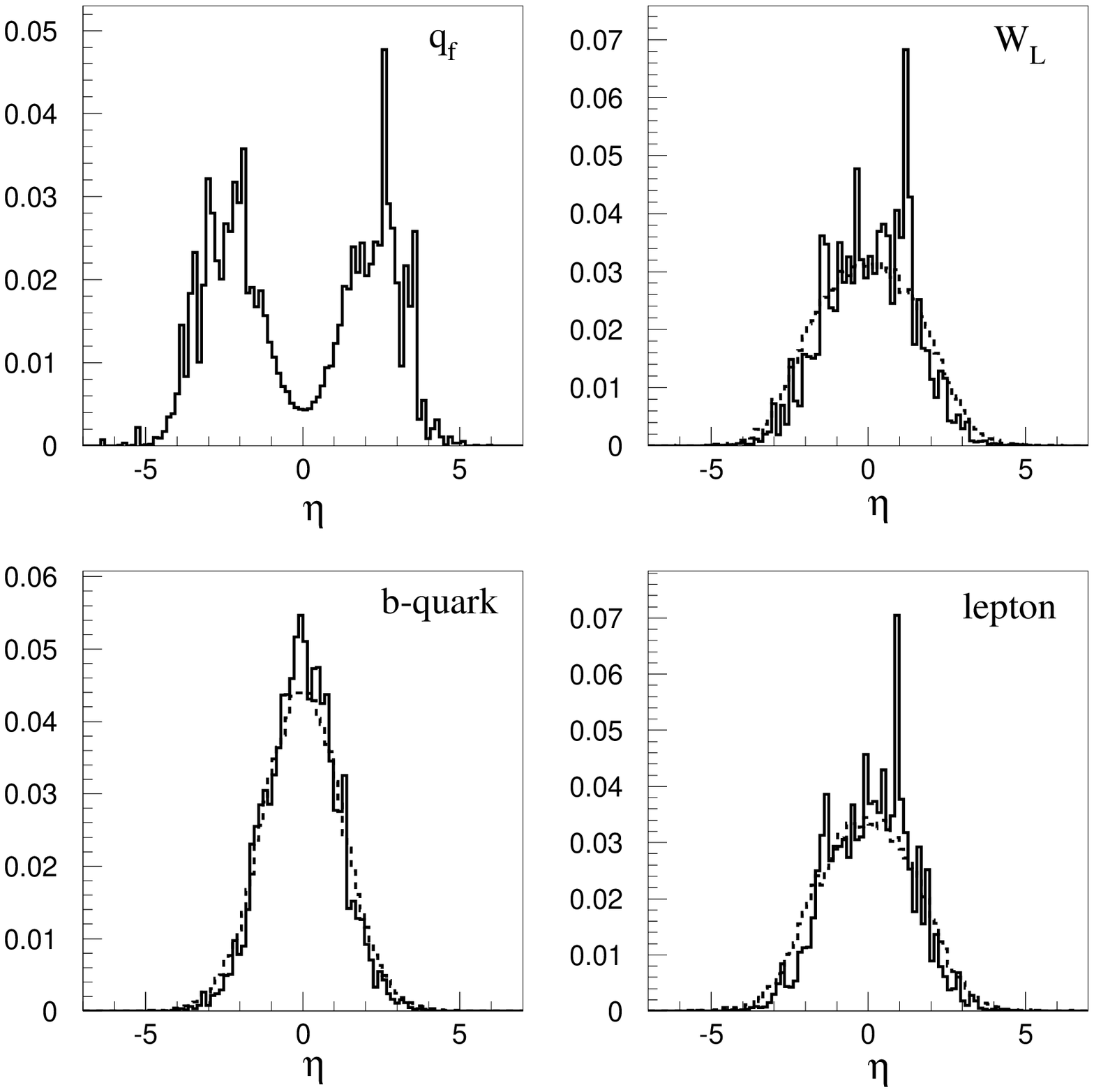,width=16cm} 
\ccaption{}{
Pseudorapidity distributions of final particles, namely,  $q_f$ quarks, 
$W_L$, $b$--quarks, and charged lepton. Solid and dashed histograms are the
results of the exact and EWA calculations, respectively. All histograms are
normalized to unit area.
}
\end{center}
\end{figure}

% Figure 6
\begin{figure}[t]
\begin{center}
\epsfig{file=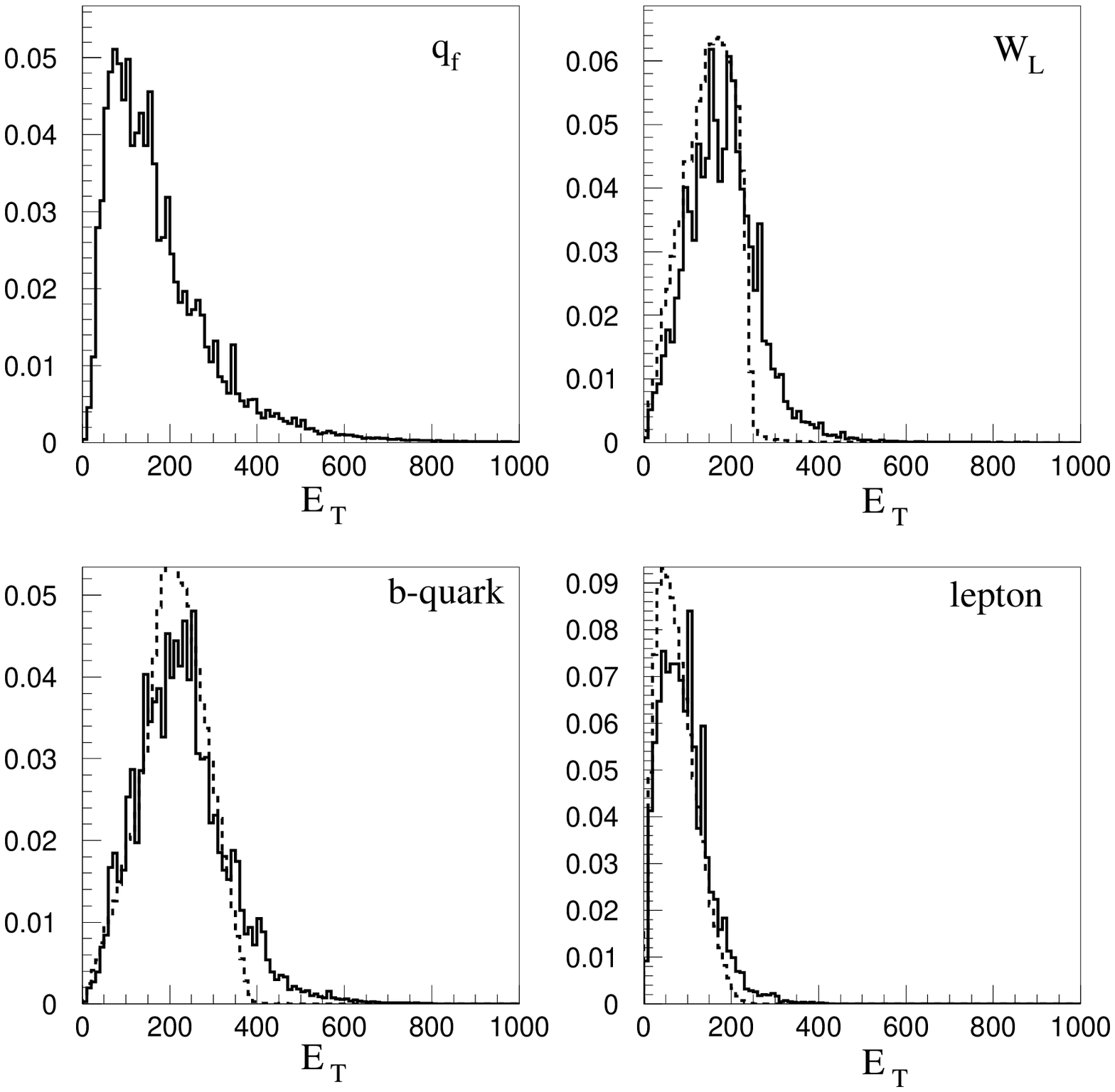,width=16cm} 
\ccaption{}{
Transverse energy distributions of final particles, namely,  $q_f$ quarks, 
$W_L$, $b$--quarks, and charged lepton.
Solid and dashed histograms are the
results of the exact and EWA calculations, respectively.
All histograms are
normalized to unit area.}
\end{center}
\end{figure}

\end{document}